# Magnetic field-modulated exciton generation in organic semiconductors: an intermolecular quantum correlation effect

B. F. Ding*, Y. Yao*, X. Y. Sun, Z. Y. Sun, X. D. Gao, Z. T. Xie, Z. J. Wang, X. M. Ding, Y. Z. Wu, X. F. Jin, C. Q. Wu[†], and X. Y. Hou[††]

*Surface Physics Laboratory, Department of Physics, Fudan University,
Shanghai 200433, People's Republic of China*

Magnetoelectroluminescence（MEL）of organic semiconductor has been experimentally tuned by adopting blended emitting layer consisting of both hole and electron transporting materials. A theoretical model considering intermolecular quantum correlation is proposed to demonstrate two fundamental issues: (1) two mechanisms, spin scattering and spin mixing, dominate the two different steps respectively in the process of the magnetic field modulated generation of exciton; (2) the hopping rate of carriers determines the intensity of MEL. Calculation successfully predicts the increase of singlet excitons in low field with little change of triplet exciton population.

*These authors contributed equally to this work.
[†] Electronic mail:   cqw@fudan.edu.cn
[††] Electronic mail:   xyhou@fudan.edu.cn

In the field of organic spintronics [1-13], magnetoelectroluminescence (MEL), i.e., magnetic field-modulated electroluminescence of organic semiconductor devices with nonmagnetic electrodes has been widely studied [5-13], because it provides a promising way to understand the spin dynamics in organic systems, such as organic semiconductors and biological systems [14]. Generally, MEL appears to enhance rapidly with magnetic field in the low field region (typically <15mT) but to saturate or even weaken in the high field region (>30mT) [5,6,10,12].

Unanswered questions in this area include the reason why smaller electric field leads to larger MEL[5,6,8,10,12,15], why insertion of an insulating layer LiF enhances MEL [10] and why minority carriers dominate the magnetic field effect in organic semiconductors [6,15-17]. These questions can be reduced to one basic issue, namely what determines the intensity of MEL in a given magnetic field. Theories of inter-system crossing (ISC)[9,10,16] or triplet-triplet annihilation (TTA) [12,13] have been proposed by some groups to qualitatively explain MEL. Both ISC and TTA could be included in a so-called spin mixing (SM) mechanism which occurs only in spin sub-levels through spin redistribution [6, 14, 18]. According to ISC or TTA, the increase of singlet excitons should occur at the expense of triplet excitons. But Reufer *et al* found that no change of singlet–triplet balance occurs, especially in low magnetic field [7]. Since MEL correlates directly with magnetic field-modulated exciton generation, it is critical to consider how magnetic field influences the process of exciton generation. Generally, exciton generation consists of three steps[19]: I. carrier injection from electrodes and transport; II. formation of a hole-electron pair at two adjacent molecules; III. formation of an exciton at one molecule by hopping. Identifying in which step SM is active and which mechanism dominates beyond this step is the essential question that must be addressed [6,18]. To this end a quantitative explanation of

MEL is indispensable.

In this letter, we describe a model formulated by considering the intermolecular correlated effect. Based on the model, the process of magnetic field-modulated step II and step III of exciton generation is quantitatively studied both experimentally and theoretically. It is found that SM only dominates magnetic field-modulated step III in high magnetic field, while spin scattering (SS), in which the redistribution of both spins and charges is involved, controls the modulation of step II in low magnetic field. By studying the MEL of the organic light emitting diode (OLED) with a blended emitting layer, it is found that hopping rate of carriers influences both SS and SM. This proves that intermolecular quantum correlated effect plays an essential role in organic MEL and suggests that the magnetic field effect in other relevant systems[3,14,16] should be reconsidered.

In present work, a special structure of OLED was designed to reveal the effect of intermolecular correlation on exciton generation. The OLED was fabricated with a blended layer of tris-(8-hydroxyquinoline) aluminum ($Alq_3$) and N,N'-bis(l-naphthyl)-N,N'-diphenyl-1,l'-biphentl-4,4'-diamine (NPB), acting as the fluorescent emitting layer, as shown in Fig. 1(a). The device area is approximately $4 \times 4$ mm$^2$. Besides, 40nm NPB and 50nm $Alq_3$ form the hole-transporting and electron-transporting layers respectively. Because of its wide band gap[20], BCP (bathocuproine) is used as an exciton-blocking layer to prevent exciton diffusion from the blended layer to the 50nm-thick $Alq_3$ layer. The anode and composite cathode are ITO and LiF/Al respectively.

Figure 1(b) shows the MEL of six representative devices with different blending volume ratios of NPB to $Alq_3$ as function of external magnetic field under constant driving current of 150μA. The MEL is expressed by $\Delta EL/EL = [(EL(B) - EL(0)]/EL(0)$, where $EL(B)$ and $EL(0)$ are the electroluminescence with and without external magnetic field $B$ respectively. Similar to the reported results[6], each curve is characterized by fast enhancement in the low field region followed by saturation in the high field region. The intensity of MEL under a given field depends strongly on the blending ratio. To confirm the relation between MEL and blending ratio, Fig. 1(c) shows $\Delta EL/EL$ at 50mT under three different driving conditions—constant driving current of 500μA, constant driving voltage of 7V, and constant initial brightness of 20cd/m$^2$. With the ratio increasing from 0% to 30%, under the condition of constant initial brightness of 20cd/m$^2$, $\Delta EL/EL$ increases from 3% to 7%. However, further increase of the ratio from 30% to 80% causes $\Delta EL/EL$ to decrease from 7% to 2%. MEL under the other two driving conditions shows similar behavior. Experiments reveal an intrinsic relation between MEL and blending ratio. Liu *et al* reported previously that electrical conductivity, which is determined by carrier mobility, reaches a minimum around the blending ratio of 30% in a blended layer of NPB and Alq$_3$[21]. The ratio of 30% agrees well with the ratio in the present experiments in which MEL reaches its maximum. Quantum mechanically, mobility of carriers is dominated by their hopping rate. The hopping rate is one of the key factors that affect the interaction among intermolecular carriers, and is intrinsically a result of correlated effect. All of the above experimental results demonstrate that a correlated effect is likely to determine the intensity of MEL.

Based on the results shown in Fig. 1, it is proposed that the Hubbard model, a typical model for correlated systems, can be used to describe such a correlated effect [22]. The Hamiltonian, including both the intermolecular transportation of electrons/holes and the interaction between electrons/holes and nuclear spins, reads

$$H = H_1 + H_2, \tag{1}$$

The first part of (1) includes hopping and Coulomb interaction of carriers.

$$H_1 = -\sum_{i,j,\sigma}(t^h_{i,i'}d^{\dagger}_{i,\sigma}d_{i',\sigma} + t^e_{j,j'}c^{\dagger}_{j,\sigma}c_{j',\sigma} + h.c.) + U\sum_{i,j}(n^h_{i\uparrow}n^h_{i\downarrow} + n^e_{j\uparrow}n^e_{j\downarrow}) + V, \qquad (2)$$

where $d^{\dagger}_{i,\sigma}$ ($d_{i,\sigma}$) and $c^{\dagger}_{j,\sigma}$ ($c_{j,\sigma}$) create (annihilate) a hole and an electron with spin $\sigma$ up (↑) or down (↓) in the $i$- and $j$-th molecule, $i'$ and $j'$ denote the neighbors of the $i$- and $j$-th molecule, $t^h_{i,i'}$ and $t^e_{j,j'}$ are the hopping rates (in unit of μeV by the uncertainty principle ΔEΔt≈ℏ/2) of hole and electron. $U$ is the Coulomb repulsive energy between two holes or two electrons with different spins at the same molecule, $n^h_{i,\sigma}$ ($\equiv d^{\dagger}_{i,\sigma}d_{i,\sigma}$) and $n^e_{j,\sigma}$ ($\equiv c^{\dagger}_{j,\sigma}c_{j,\sigma}$) the corresponding hole and electron number operators, and $V$ the attractive interaction between hole and electron at the same molecule.

The second part of (1) shows the effect of an external magnetic field and hyperfine interactions,

$$H_2 = g\mu_B \sum_{i,j} \left[ (\vec{B}_{ext} + \vec{B}_{hyp,i}) \cdot \vec{S}^h_i + (\vec{B}_{ext} + \vec{B}_{hyp,j}) \cdot \vec{S}^e_j \right], \qquad (3)$$

where g=2.0 for organic materials[7], $\mu_B$ is Bohr magneton, $\vec{B}_{ext}$ is the external magnetic field chosen along the z-direction, and $\vec{B}_{hyp,i(j)}$ is the effective nuclear magnetic field of $i$-th ($j$-th) molecule that will be treated classically. For example, for holes,

$$\vec{B}^h_{hyp,i} \cdot \vec{S}^h_i = B^h_{hyp,i}\cos\theta^h_i(d^{\dagger}_{i,\uparrow}d_{i,\uparrow} - d^{\dagger}_{i,\downarrow}d_{i,\downarrow}) + B^h_{hyp,i}\sin\theta^h_i(d^{\dagger}_{i,\uparrow}d_{i,\uparrow} + d^{\dagger}_{i,\downarrow}d_{i,\downarrow}), \qquad (4)$$

where $\theta^h_i$ is the angle between $\vec{B}_{hyp,i}$ and hole spin.

In this work, we focus on the intrinsic features of the organic semiconductor. The model is a two-step one including only step II and step III of exciton generation mentioned above. Because step II and step III are different, they are treated separately. Generally, hopping rates $t^h_{i,i'}$ and $t^e_{j,j'}$ vary from molecule to molecule due to the disorder of organic materials. $t^h$ and $t^e$ are estimated to be between 1μeV and 100μeV[23], depending on the blending ratio. U is taken as 0.1eV[24], which is large enough to prevent double occupancy of the same polarity of carriers at one molecule, yet spin exchange between two molecules still operates when many excitons interacting with each other in step III[22] are considered. Because of the random orientation of molecular $\vec{B}_{hyp}$, it is necessary to calculate the statistical average for the angle of nuclear spins $\theta(\in(0,\pi))$. Hyperfine interaction $g\mu_B B_{hyp,i(j)}$ is set to 0.5μeV[24,25]. Thus an external magnetic field about five times higher than the hyperfine interaction (>25mT) can be considered as a high field. Then, given an initial state that electron and hole reside on different molecules, the average generation probability of both singlet and triplet excitons can be calculated over the time range 0 to 10ns, covering the mean lifetime of excitons.

Figure 2 shows results of the present simulation, compared to the corresponding experimental measurements. Without loss of generality, all calculations have been normalized with the result of pure Alq$_3$ structure. Each curve contains two parts corresponding to magnetic field-modulated step II and step III respectively, as marked in the figure. Calculated $\Delta EL/EL$, shown as solid lines, is consistent with the experimental data. It is seen that for step II the calculated $\Delta EL/EL$ increases quickly in low magnetic field, showing close agreement with the experimental results. For step III, the calculated $\Delta EL/EL$ saturates and reproduces the experimental data above 25mT. Two insets in Fig. 2 show the hopping rate parameters for magnetic field-modulated step II and magnetic field-modulated step III. Hopping rates in step II are taken to be one order smaller than those used in step III, since in step III the stronger Coulomb interaction between electron and hole increases the hopping rates[27]. It is found that the calculated $\Delta EL/EL$ increases monotonically with decreasing hopping rate in both steps, behaving in the same way as experimentally observed. It should be emphasized that present work only focuses on two regions of MEL, the low and the high field region. The transition region will be considered in future. These results confirm that the correlated effect, which is controlled by the intermolecular hopping rate, is essential for MEL and that lowering the hopping rate can enhance the MEL.

To understand the physics of the two-step model, a possible illustration is given in Fig. 3, showing schematically how step II and III are influenced in different magnetic field regions. Magnetic field-modulated step II is shown in Fig. 3(a) and Fig 3(b), demonstrating how a low external magnetic field influences the SS, and thus increases the generation probability of hole-electron pairs. If no external magnetic field $\vec{B}_{ext}$ exists, the intensity value of $\vec{B}_{hyp}$, the effective nuclear magnetic field of a molecule, is identical for each molecule, as shown by black arrows in Fig. 3(a). This means that the intensity distribution of $B_{hyp}$ is uniform in space. However, for a given low $\vec{B}_{ext}$, the total $\vec{B}_{tot}$ of $\vec{B}_{ext}$ and $\vec{B}_{hyp}$ is not identical due to the random orientation of $\vec{B}_{hyp}$, resulting in non-uniform intensity distribution of $B_{tot}$, as shown by red arrows in Fig. 3(a). If $B_{ext} \gg B_{hyp}$, all $\vec{B}_{tot}$ are almost the same as $\vec{B}_{ext}$ and the intensity distribution of $\vec{B}_{tot}$ should be uniform. The non-uniform $\vec{B}_{tot}$ acts like disorder in energy distribution for the carrier spins, strengthening their scattering and resulting in enhancement of the formation probability of hole-electron pairs. The dependence of spin scattering on $B_{ext}$ is very similar to that in the case of dilute magnetic semiconductors, where the spin scattering between carrier and magnetic atom is proportional to x(1-x), where x is the density of magnetic atoms[28]. In Fig. 3(b), the carrier capture cross section is used to describe the spin scattering. Without $\vec{B}_{ext}$, the intensity distribution of $\vec{B}_{hyp}$ is uniform. The carrier capture cross section for a molecule is small, as shown in the left part of Fig. 3(b). Low $\vec{B}_{ext}$ will cause non-uniform intensity distribution of $\vec{B}_{tot}$, and thus enlarge the carrier capture cross section, as shown in the right part of Fig. 3(b). When $B_{ext} \gg B_{hyp}$, the intensity distribution of $\vec{B}_{tot}$ is uniform again and the carrier capture cross section will decrease to the value in the case of $B_{ext}=0$. Magnetic field-modulated step III is shown in Fig. 3(c) and Fig. 3(d), demonstrating how a high external magnetic field suppresses the spin flip, and thus increases the ratio of singlet to triplet excitons. In the case of $B_{ext}=0$, as shown in the left part of Fig. 3(d), a hole and an electron are located at two adjacent molecules to form a singlet hole-electron pair. When the electron hops to the adjacent molecule where the hole resides, the electron spin will be modulated by both $\vec{B}_{hyp}^{h}$ and $\vec{B}_{hyp}^{e}$ of the two molecules where hole and electron reside respectively. Due to their random orientation, $\vec{B}_{hyp}^{h}$ and $\vec{B}_{hyp}^{e}$ point to different directions. In such case, $\vec{B}_{hyp}^{e}$ provides the transversal component to $\vec{B}_{hyp}^{h}$. It is well known that the transversal component makes spin flip feasible during electron hopping[28]. Therefore the singlet hole-electron pair can undergo transition to a triplet exciton, as shown in the left part of Fig 3(d). This is indeed the mechanism of SM caused by random nuclear field[29]. However, a high $B_{ext}$ can align $B_{tot}$ towards $B_{ext}$, therefore $B_{tot}$ would no longer be random as $B_{hyp}$, as shown by red arrows in Fig. 3(c). The right part of Fig. 3(d) shows when $B_{ext} \gg B_{hyp}$, compared to the large longitudinal component of $\vec{B}_{tot}$ the transversal component becomes insignificant. In this case spin flip will hardly occur. Therefore the inhibition of SM effectively increases the number of singlet excitons. MEL in different ranges of magnetic field is dominated by different mechanisms - SS and SM. They dominate the magnetic field-modulated step II and step III respectively.

From the two-step model, the intrinsic property of the mechanism of SS is that increase of singlet excitons is not obtained solely at the expense of triplet excitons. Fig. 4(a) shows the theoretical result concerning the variation of singlet and triplet excitons with low magnetic field. It can be seen that the influence of magnetic field on singlet excitons is much stronger than that on triplet excitons. Experimentally, Reufer *et al* have examined whether an external magnetic field could change the singlet/triplet ratio by using a phosphorescent hydrocarbon polymer with simultaneous emission of fluorescence and phosphorescence, but the method could hardly prevent the internal energy transfer between singlet and triplet excitons[7,18]. In order to eliminate the internal energy transfer, a special structure of OLED, as schematically shown in Fig. 4(b), was adopted, which was similar to the structure reported before for studying the phosphorescence of phosphorescent dye 2,3,7,8,12,13,17,18-octaethyl-21H,23H-porphine platinum(II) (PtOEP)[30]. In Fig. 4(b), C6 stands for Coumarine 540. Since C6 (PtOEP) is a fluorescent (phosphrescent) dye with efficient Förster (Dexter) energy transfer from singlet (triplet) exciton of $Alq_3$ to singlet (triplet) exciton of C6 (PtOEP), C6 (PtOEP) doped $Alq_3$ layer with volume doping ratio of 1.0% (4.0%) acts as the singlet (triplet) exciton recombination zone. The singlet exciton recombination zone and the triplet exciton recombination zone are separated by a 20nm $Alq_3$ layer. 20nm is sufficiently thick to avoid internal energy transfer between the two recombination zones[30]. Thus the variation of singlet and triplet excitons in magnetic field should be observed simultaneously. Fig. 4(c) shows the MEL of the device under a field of 6mT. The peaks of 510nm and 652nm are emitted from the singlet exciton recombination zone and the triplet exciton recombination zone respectively. The spectra adjacent to the two peaks are amplified and shown in the left and right insets. With field of 6mT, $\Delta EL/EL$ around 510nm is calculated to be about 2.5%, while $\Delta EL/EL$ around 652nm is almost zero. Fig. 4(d) shows the experimental results of $\Delta EL/EL$ for singlet excitons and triplet excitons in the low field range. Increasing the field from 1mT to 6mT, $\Delta EL/EL$ for singlet excitons increases from 0.2% to 2.5%, while $\Delta EL/EL$ for triplet excitons is nearly zero over this range. Similar to the results shown in Fig. 4(a) these experimental results also indicate that the influence of the low magnetic field on singlet excitons is much stronger than that on triplet excitons. Some difference exists between the theoretical result shown in Fig. 4(a) and the experimental result shown in Fig. 4(d) for the triplet excitons. It indicates more physical processes should be taken into consideration in the model. Furthermore, it must be mentioned that the structure shown in Fig. 4(b) can not be used experimentally to observe the behavior of triplet excitons in high magnetic field, since high field influences many unclear effects in PtOEP, such as spin-orbit interaction and spin-spin interaction, which in turn influence the experimental result.

In conclusion, we report that MEL strongly depends on the blending ratio of NPB to $Alq_3$. The maximum MEL occurs at the blending ratio of 30%, which corresponds to the minimum conduction of blended layer. Compared to the theoretical result, we confirm that the intermolecular correlations play the essential role in MEL. In addition, to confirm the result as predicted by the proposed model, we adopt the special structure of OLED to avoid internal energy transfer between singlet and triplet exciton，so as to observe them simultaneously and individually. The increase of singlet excitons in low field with little change of triplet exciton population exceeds all experimental and theoretical results reported before.

## Acknowledgements

This work is supported by the National Natural Science Foundation of China (Grant No 10621063) and Shanghai Science and Technology Commission (Grant No 08JC1402300) and the MST of China (Grant No. 2009CB929200). YY and CQW are also supported by the EC Project of OFSPIN (Grant No. NMP3-CT-2006-033370). We would like to thank Edward Obbard for his assistance in proofreading the manuscript.

# Figure legends

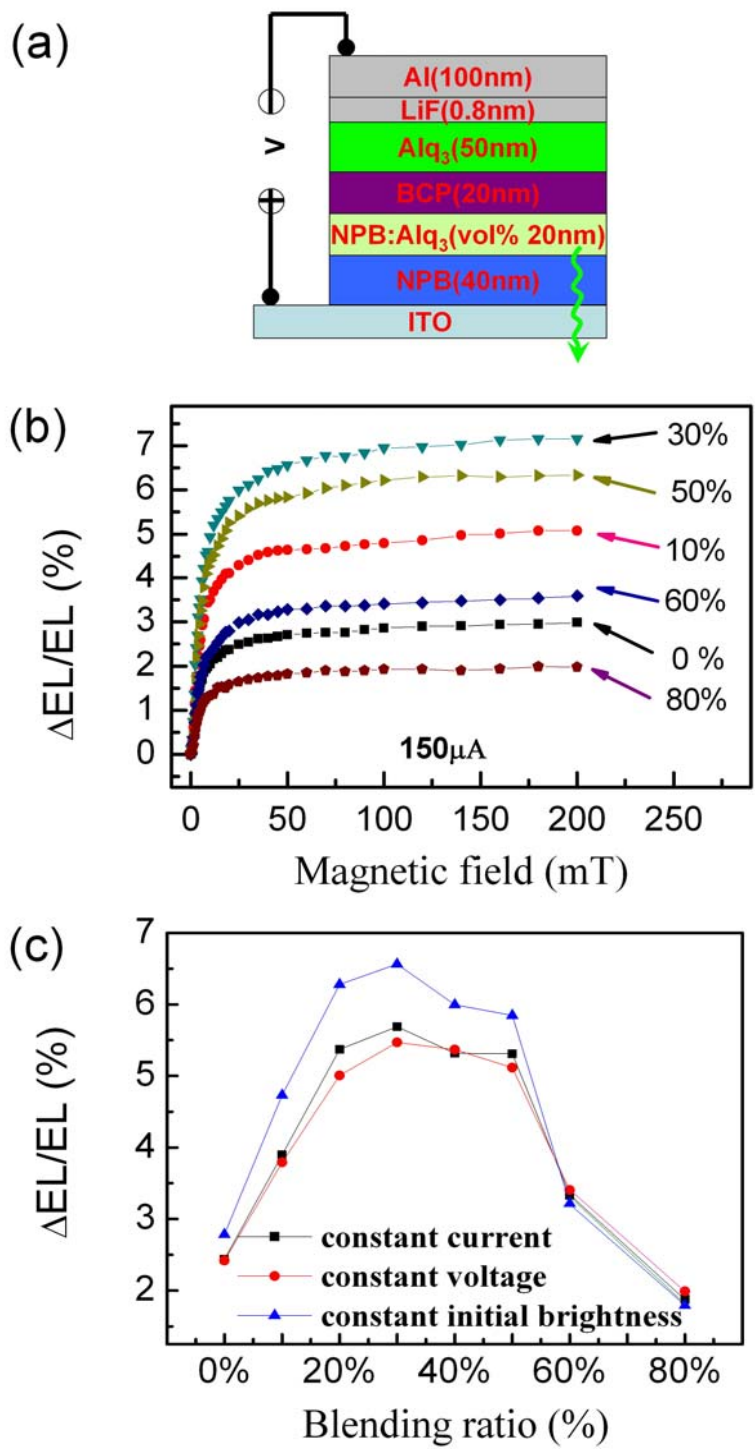

**FIG. 1 |**
MEL versus magnetic field and blending ratios of NPB to Alq3 of blending layer. (a) Schematic structure of blending-layer-based OLED. (b) MEL of six devices versus external magnetic field. For all of the devices, the measurement is under constant current driving of 150μA. (c) MEL in field of 50mT versus blending ratios for three different driving conditions (black squares: constant current of 500μA; red circles: constant voltage of 7V; blue triangles: constant initial brightness of

20cd/m$^2$).

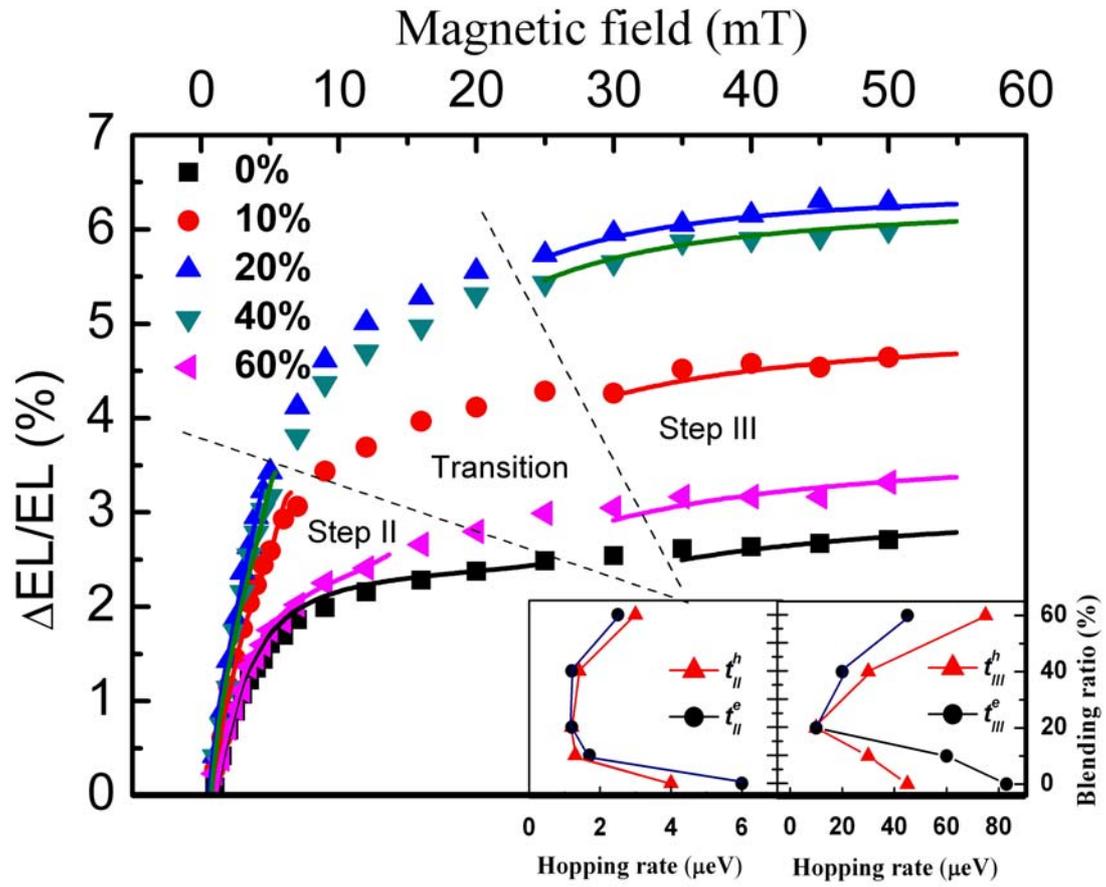

**FIG. 2 |**

Results of calculation. Dots are experimental data with five representative blending ratios, while the solid lines show the relative results of the present two-step model calculation. Two sets of solid lines in low field and high field are results of calculation for the magnetic-field-modulated step II and step III of the exciton generation. Hopping rate parameters are shown in the insets: $t_{II}^{h}$ and $t_{II}^{e}$ ($t_{III}^{h}$ and $t_{III}^{e}$) are hopping rate of hole and electron respectively for magnetic-field-modulated step II (step III).

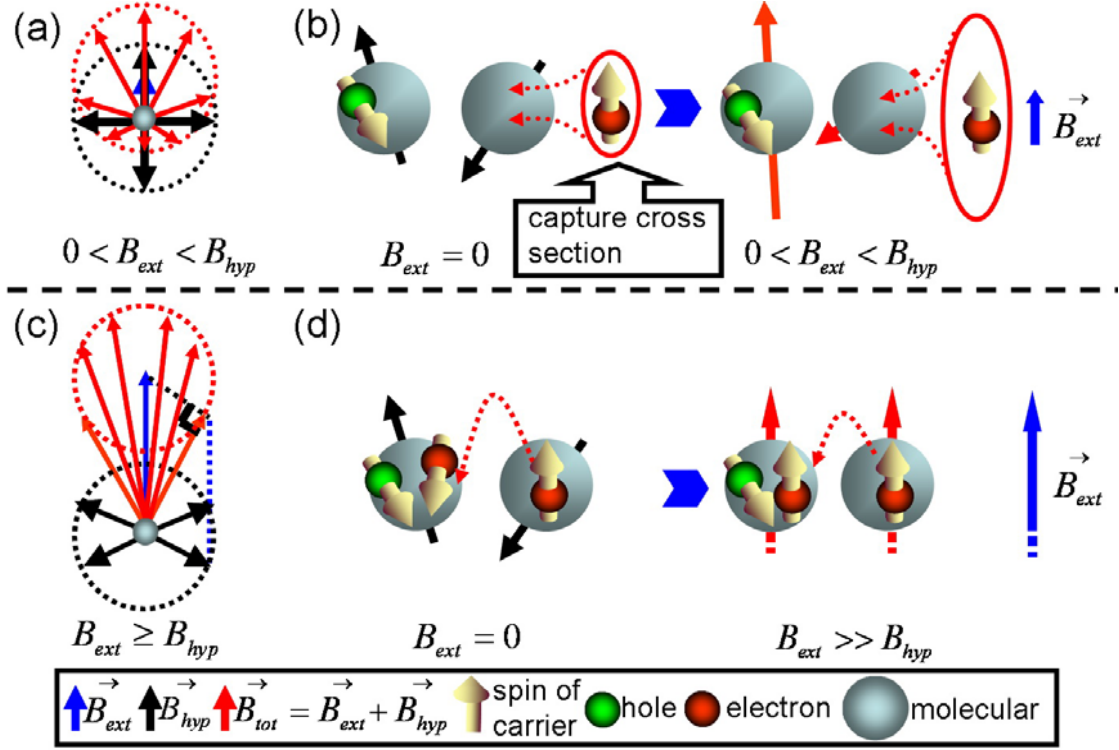

**FIG. 3 |**

Schematic diagram of the influence of external magnetic field in different ranges on the exciton generation. Blue arrow/black arrow represents external magnetic field $\vec{B}_{ext}$/effective nuclear magnetic field $\vec{B}_{hyp}$, and red arrow stands for $\vec{B}_{tot}$ (the resultant field of $\vec{B}_{ext}$ and $\vec{B}_{hyp}$). (a) Low $\vec{B}_{ext}$ changes the uniform distribution of intensity of $\vec{B}_{hyp}$ into non-uniform distribution. (b) Left part indicates that without $\vec{B}_{ext}$ carrier capture cross section is small. Right part indicates that the non-uniform distribution caused by low $\vec{B}_{ext}$ increases carrier capture cross section, and thus increases the formation probability of hole-electron pairs. (c) High $\vec{B}_{ext}$ can align random orientation of $\vec{B}_{hyp}$ towards the same orientation. (d) Left part shows how singlet hole-electron pair transits to triplet exciton through spin flip without $\vec{B}_{ext}$. Right part shows when $\vec{B}_{ext} \gg \vec{B}_{hyp}$, the spin flip is ineffective, and the singlet hole-electron pair can only transit to singlet exciton.

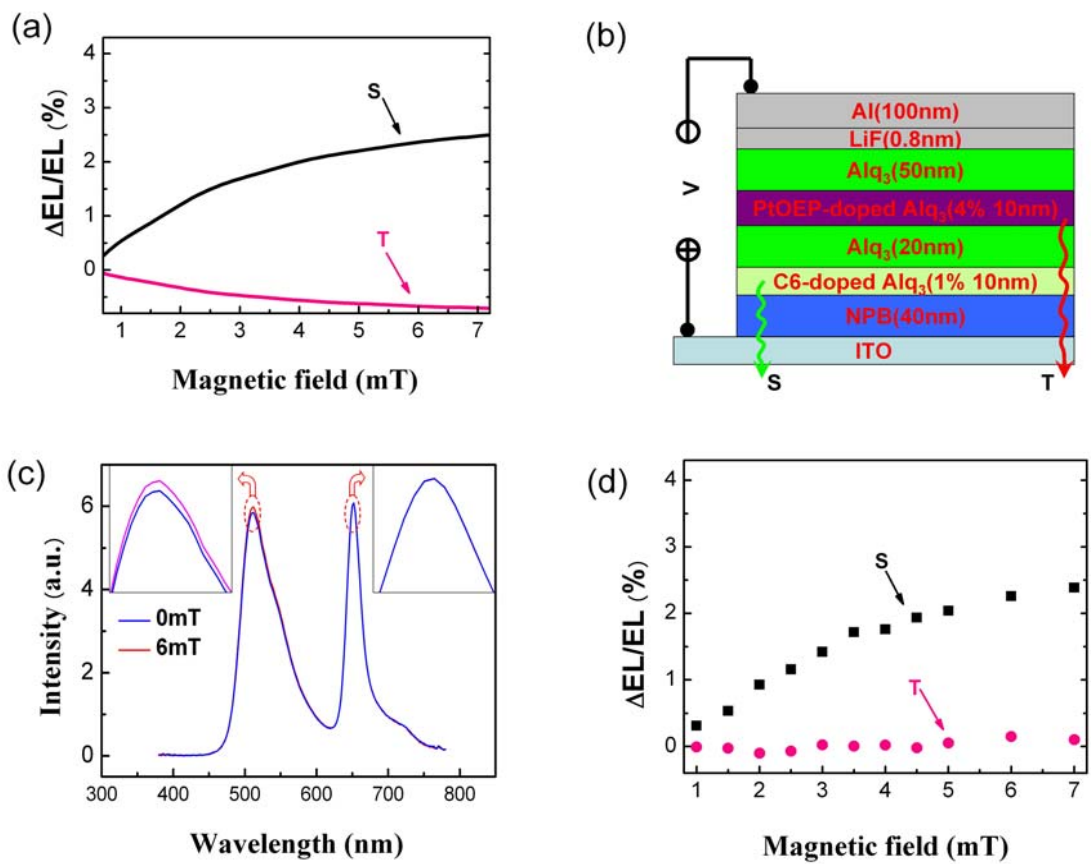

**FIG. 4|**

Behaviors of singlet and triplet excitons in the range of low magnetic field where spin scattering dominates. (a) Theoretical calculation of the behavior of singlet and triplet excitons: singlet excitons are more sensitive to magnetic field than triplet excitons. Solid black and red lines are for singlet and triplet excitons respectively. (b) Schematic architecture of devices consisting of a layer doped with fluorescent dye C6 for detecting and removing singlet excitons and another layer doped with a phosphorescent dye for detecting triplet excitons. (c) Spectra obtained from the device with structure shown in (b): blue spectrum is measured without external field; red spectrum is measured under 6mT field. Left and right insets show amplified curves around two peaks. (d) Experimental results for behaviors of singlet and triplet excitons in low magnetic field.